\documentclass[10pt, a4paper]{article}
\usepackage[a4paper, margin=1.8cm]{geometry}
\usepackage{graphicx}
\usepackage{subcaption}

\usepackage{amsmath, amsfonts, amssymb}
\usepackage{booktabs}
\usepackage{cite}
\usepackage{color}
\usepackage{xcolor}
\usepackage{hyperref}

\title{\textbf{\Large Agentic Limit Order Books: Phase Transitions and Market Impact}}

\author{
    \textbf{Jan Rosenzweig}\\
}

\begin{document}

\maketitle

\begin{abstract}
We investigate the systemic macroscopic dynamics emerging from Limit Order Books (LOBs) populated exclusively by autonomous reinforcement-learning agentic traders. By formalizing agent interactions within a microscopic order-matching engine, we examine two fundamental quantitative phenomena: equilibrium phase transitions in order flow regime shifts, and the structural dynamics of market impact. We show that agentic LOBs exhibit distinct phase boundaries separating orderly price discovery from hyper-volatile cascade states, governed by critical thresholds in the number of agents and observable market depth. Furthermore, we demonstrate that market impact under agentic liquidity provision deviates from classical square-root dynamics, exhibiting distinct dissipative, balanced, and non-dissipative regimes under non-linear feedback loops.
\end{abstract}

\section{Introduction}
Modern quantitative finance is undergoing a fundamental structural paradigm shift. The historical transition from manual execution to rule-based algorithms (such as TWAP, VWAP etc) is now giving way to fully autonomous, agentic trading architectures. Driven by advances in reinforcement learning and large-scale reasoning systems, contemporary software agents do not simply execute static instructions; they dynamically interpret order book states, infer competitor intent, adapt execution horizons, and adjust inventory limits in real time.

While agentic workflows promise enhanced micro-structural efficiency and superior adaptive execution for individual market participants, their aggregate behavior in closed-loop ecosystems presents unprecedented systemic challenges. As emphasized by Farmer \& Foley \cite{farmer2009}, classical equilibrium models fail to capture emergent macro-level dynamics in financial networks, making agent-based modeling essential. When limit order books (LOBs) are populated predominantly by adaptive agents, it is unclear whether standard market microstructure assumptions regarding order arrival rates, liquidity provision, and price impact still stand, or whether they break down and how \cite{kyle1985, hasbrouck2007}. The strategic co-adaptation of these agents creates intricate non-linear feedback loops, giving rise to emergent macro-level phenomena that cannot be predicted by analyzing single-agent behavior in isolation.

Our main concern, however, is the potential use of agents in simulation. By populating an LOB with autonomous agents, we get a reactive environment against which to test trading strategies. Unlike in traditional tick-level simulations, which rely on impact models to predict how the other traders would or could react to trading and order flow from a given strategy, an agentic order book is by construction a reactive environment, in which such reactions are played out as they would be in live trading. 

To use a chess analogy, if trading against a historical simulation is analogous to studying historical chess games, trading against agents is then analogous to playing against a chess engine.

To better understand the statistical physics and quantitative market microstructure \cite{cont2010} of such systems, this manuscript presents a formal mathematical and computational framework for analyzing aggregate dynamics in \textit{Agentic Limit Order Books} (A-LOBs). Specifically, we focus on two core financial phenomena:
\begin{enumerate}
    \item \textbf{Phase Transitions and Market Stability:} We map the parameter space governing agent population sizes and market depth to establish critical phase boundaries separating stable, efficient price discovery regimes from frozen order books and catastrophic liquidity collapse.
    \item \textbf{Market Impact and Dissipation Dynamics:} We examine how transient and permanent market impact functions behave when liquidity providers are themselves dynamic agents, detailing the transition from dissipative (absorptive) markets to non-dissipative (cascade-inducing) regimes.
\end{enumerate}

Understanding these dynamics is paramount not only for quantitative traders seeking to optimize execution strategies in agent-dominated markets, but also for risk managers and market regulators charged with maintaining orderly price discovery amidst automated cascade risks.

\section{Model Setup \& Agent Topology}
Consider a discrete-time continuous double auction populated by $N$ autonomous agents, partitioned into three functional classes: Market Agents ($\mathcal{M}$), Scalpers ($\mathcal{S}$), and Liquidity-Providing agents ($\mathcal{L}$). 

Liquidity-Providing Agents operate with a fixed depth horizon $d$ and quantity limit $L$, observing the Order Book up to $\pm d$ ticks from the best bid and ask. They are trained to maximize trading PnL within a fixed time horizon by issuing and modifying limit orders, while employing market orders primarily to unwind inventory when breaching the holding threshold $\pm L$.

Scalpers are topologically analogous to Liquidity-Providing Agents but operate with a constrained market depth of $d=2$, observing only the touch and next-to-touch levels on both sides of the book.

Market Agents act as stochastic liquidity takers, issuing random market buy and sell orders at a fixed frequency with sizes bounded by a maximum limit.

The limit order book state at discrete time step $t$ is represented by the array $S_t = (S_t^b, S_t^a, Q_t^b, Q_t^a)$, denoting bid/ask prices and quantities respectively. Each agent $i \in \{1, \dots, N\}$ reacts to book updates, executed trades, own fills, and timed callbacks by generating order actions:
\begin{equation}
a_{i,t} \in \{\text{Limit Buy}, \text{Limit Sell}, \text{Market Buy}, \text{Market Sell}, \text{Modify/Cancel}\}
\end{equation}

The mid-price evolution follows the standard relation:
\begin{equation}
    S_t = \frac{S_t^{b,1} + S_t^{a,1}}{2}
\end{equation}
where $S_t^{b,1}$ and $S_t^{a,1}$ denote the best bid and ask prices, respectively.

\section{Phase Transitions in Agentic LOBs}
The interaction of adaptive agents in a shared order book naturally gives rise to collective phase behavior. Given that altering the number of Market Agents is mathematically equivalent to scaling their execution frequency, we fix the number of Market Agents to $|\mathcal{M}| = 1$.

We analyze isotropic order books where all Liquidity Providers share identical parameters $(d, L)$, and all Scalpers operate with an identical limit $L$. We evaluate system dynamics as a function of the number of Liquidity Providers $n = |\mathcal{L}|$ and their observable Market Depth $d$, while scaling the Scalper population proportionally as $\max(3, \lfloor n/3 \rfloor)$. The floor of three Scalpers is strictly required to prevent instantaneous structural collapse; below this threshold, 0--2 Scalpers fail to consistently form the touch, depriving Liquidity Providers of reference prices.

Formally, this agentic system maps directly onto an open statistical mechanical system:
\begin{itemize}
    \item The volume executed by Market Agents serves as the driving \textbf{temperature} ($T$).
    \item The number of Liquidity Providers $n$ represents the system's \textbf{thermal capacity} ($C_v$).
    \item The observable Market Depth $d$ maps to system \textbf{volume} ($V$), or inversely, its effective thermodynamic \textbf{pressure} ($P$).
\end{itemize}

Varying $n$ and $d$ thus controls the ratio of driving temperature to thermal capacity alongside the effective system pressure.

Each order book path is instantiated with identical initial conditions and allowed to iterate for 100,000 timer callbacks across 10,000 Monte Carlo paths per parameter configuration. We track two macroscopic observables:
\begin{enumerate}
    \item Mid-price normal volatility $\sigma$, measured in ticks.
    \item Collapse probability $p$, defined as the empirical likelihood that the order book completely empties (i.e., zero liquidity on at least one side) before path completion.
\end{enumerate}

\subsection{Regime Shifts and Critical Thresholds}
As the agent count $n$ and market depth $d$ vary, the system undergoes sharp first-order phase transitions separating three distinct macroeconomic regimes:
\begin{enumerate}
    \item \textbf{Vaporized / Collapsed Phase ($n < n_c$):} Insufficient liquidity provision to absorb external order flow, driving the order book to extinguish in finite time ($p > 90\%$). Such abrupt regime shifts closely reflect critical power-law fluctuation behaviors found across financial markets \cite{gabaix2003}.
    \item \textbf{Continuous Liquidity Phase ($n > n_c, d > d_c(n)$):} Stable, functional markets operating with robust liquidity, finite non-zero volatility, and efficient price discovery.
    \item \textbf{Frozen Phase ($d < d_c(n)$):} Excessive agent density compressed into inadequate market depth. Price volatility vanishes ($\sigma \ll 1$ tick) as agents lock into tight queue positions, stifling price discovery.
\end{enumerate}

\begin{figure}[h!]
    \centering
    \includegraphics[height=0.15\textheight]{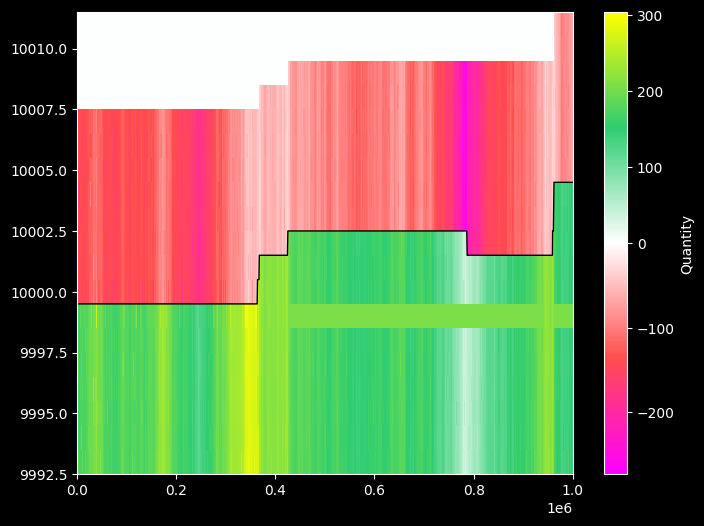} 
    \includegraphics[height=0.15\textheight]{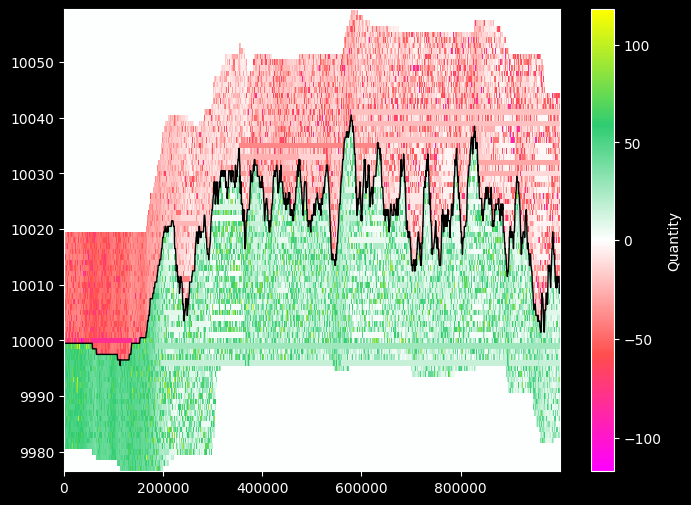} 
    \includegraphics[height=0.15\textheight]{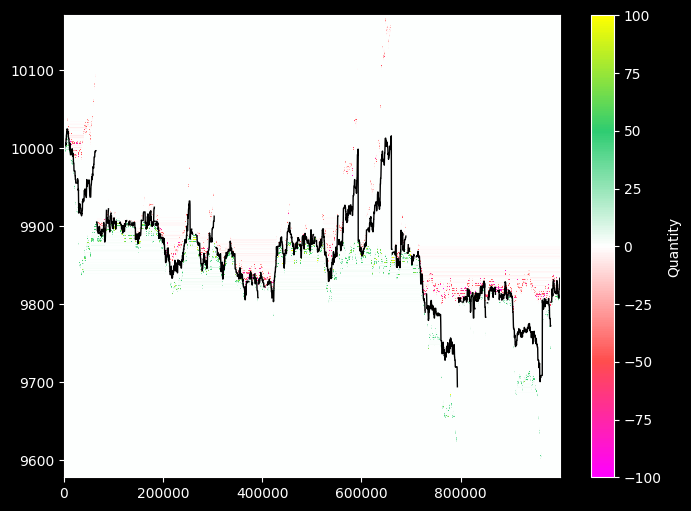} 
    \caption{Evolution of the Order Book in the Solid (left), Liquid (centre) and Vapour (right) Phases.\\
    Positive quantities denote Bids, negative quantities denote Asks, Mid Price is shown in black.}
    \label{fig:evolution}
\end{figure}

Typical Order Book and Mid Price trajectories for all three phases are shown in Figure \ref{fig:evolution}.

The physical regimes and corresponding financial market states are summarized in Table \ref{tab:regimes}.

\begin{figure}[t]
    \centering
    \includegraphics[width=0.48\linewidth]{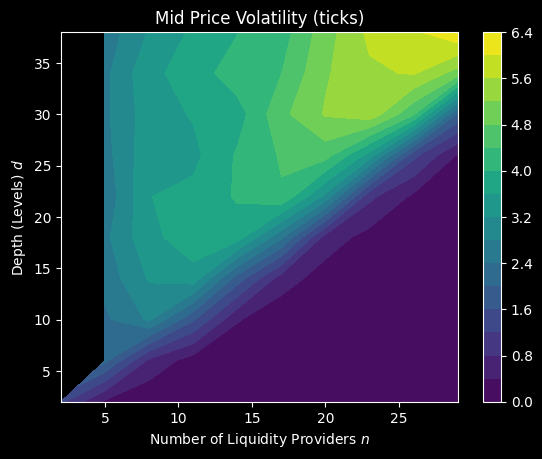}
    \includegraphics[width=0.48\linewidth]{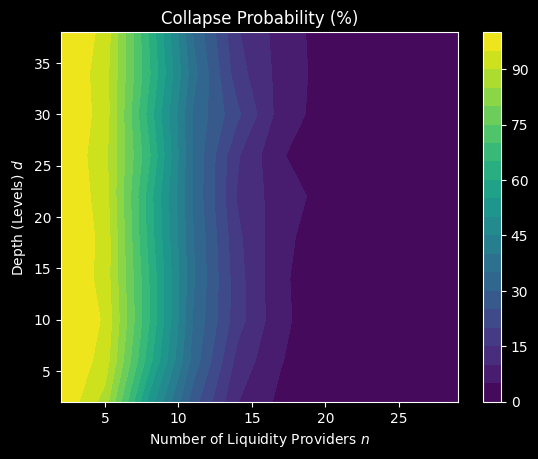}
    \caption{Mid-price volatility $\sigma$ and probability of order book collapse $p$ across varying Liquidity Provider count $n$ and Market Depth $d$. System configuration: 1 Market Agent (Max Qty = 50, frequency = 1 trade/callback), $n$ Liquidity Providers (Max Qty = 50), and $\max(3, \lfloor n/3 \rfloor)$ Scalpers (Max Qty = 10).}
    \label{fig:phases}
\end{figure}

\begin{table}[h]
\centering
\caption{System Regimes in Agentic Limit Order Books}
\vspace{2mm}
\label{tab:regimes}
\begin{tabular}{@{}lll@{}}
\toprule
\textbf{Phase} & \textbf{Control Parameters} ($n,d$) & \textbf{Market State} \\ \midrule
Collapsed Order Book & $n < n_c$ & Cascade / Breakdown \\
Continuous Liquidity  & $n > n_c,\ d > d_c(n)$ & Stable / Efficient Price Discovery \\
Frozen Order Book & $d < d_c(n)$ & Illiquid / Stalled Price Discovery \\ \bottomrule
\end{tabular}
\end{table}

Empirically, our simulation results show that the boundary separating the Collapsed and Continuous phases is fixed in $n$ and invariant to depth $d$, whereas the boundary separating Continuous and Frozen regimes scales linearly with agent density:
\begin{equation}
    d_c \propto n, \quad n_c = \text{const.}
    \label{phaseBoundary}
\end{equation}

These scaling relations are intuitively clear. Adding a liquidity provider requires proportional expansion in spatial depth $d$ to maintain queue flexibility. Conversely, once minimum operational capacity $n_c$ is met, zero-mean non-directional flow cannot systematically deplete the book, rendering depth parameter $d$ secondary to overall structural stability.

It is important to note that the phase behaviour illustrated in Figure \ref{fig:phases} depends on the aggressive flow being drift-free. If the aggressive flow is significantly biased, it will eventually fill a finite number of Liquidity Providers for any fixed $n$, resulting in the system reaching the Collapsed Phase in finite time for $n>n_{c}$.

\section{Market Impact Dynamics}
To evaluate market impact, we inject a single aggressive liquidity-taking order into the system. The book is initialized and run for $t_Q = 10$ timer callbacks, at which point an aggressive buy order of size $Q$ is executed. The simulation is then continued for the remainder of the path. 

Impact is measured via the $z$-score comparing impacted paths ($Q=200$) against non-impacted control paths ($Q=0$),
\begin{equation}
    z(t) = \frac{\mu_{I}(t) - \mu_{N}(t)}{\sqrt{\sigma_{I}(t)\sigma_{N}(t)}}
    \label{z}
\end{equation}
where $\mu_{I}(t), \sigma_{I}(t), \mu_{N}(t), \sigma_{N}(t)$ are the ensemble mean and standard deviation of the impacted paths and non-impacted paths respectively, evaluated at time $t$.

\subsection{Market Impact in the Continuous Liquidity Phase}
\begin{figure}[h!]
    \centering
    \begin{subfigure}[b]{0.31\textwidth}
        \includegraphics[width=\linewidth]{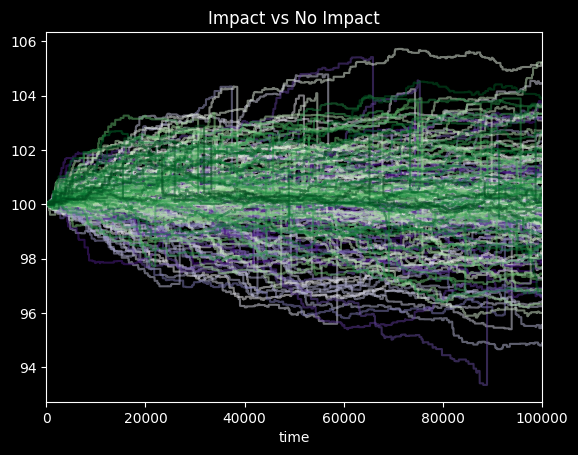}
        \includegraphics[width=\linewidth,height=0.15\textheight]{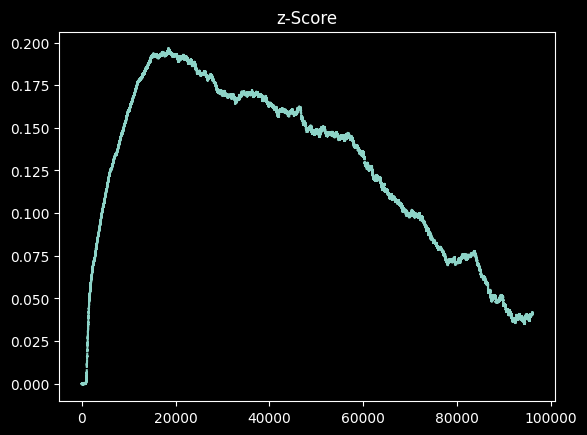}
        \caption{Dissipative; $n=11$, $d=38$; $\sigma = 3.232$ ticks}
        \label{fig:liquid:dissipative}
    \end{subfigure}
    \hfill
    \begin{subfigure}[b]{0.31\textwidth}
        \includegraphics[width=\linewidth]{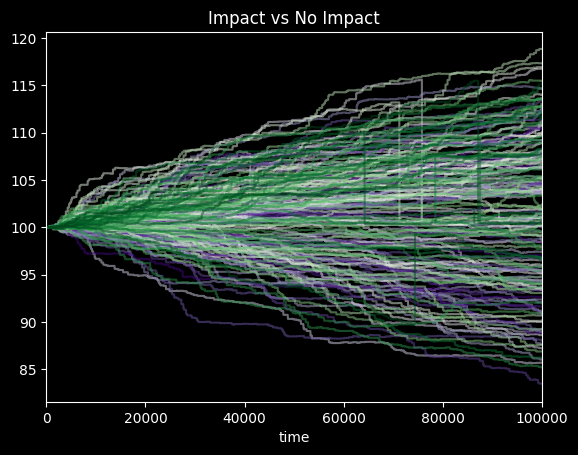}
        \includegraphics[width=\linewidth,height=0.15\textheight]{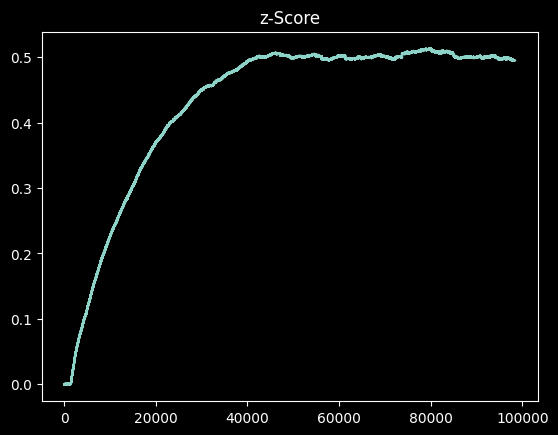}
        \caption{Balanced; $n=20$, $d=38$; $\sigma = 4.2105$ ticks}
        \label{fig:liquid:balanced}
    \end{subfigure}
    \hfill
    \begin{subfigure}[b]{0.31\textwidth}
        \includegraphics[width=\linewidth]{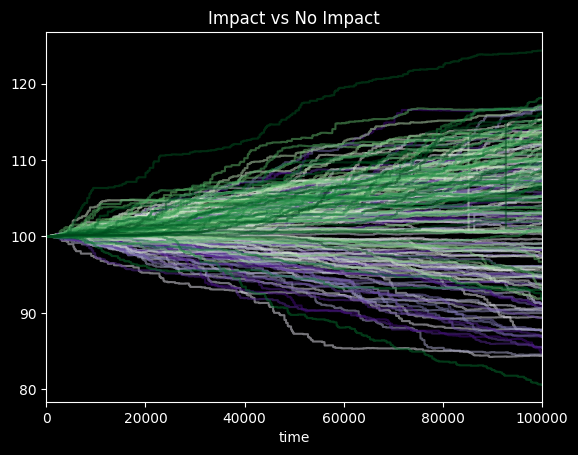}
        \includegraphics[width=\linewidth,height=0.15\textheight]{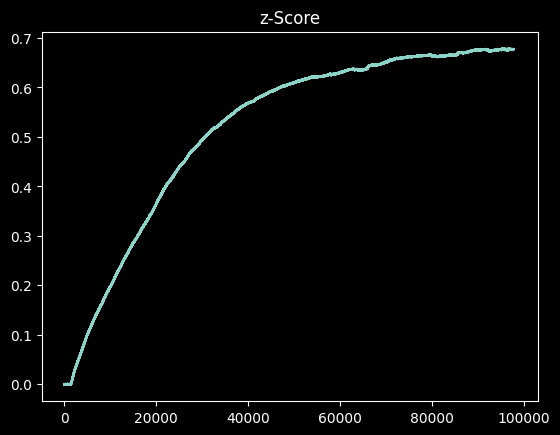}
        \caption{Non-dissipative; $n=23$, $d=30$; $\sigma = 5.529$ ticks}
        \label{fig:liquid:nondissipative}        
    \end{subfigure}
    \caption{Impact and dissipation in the Continuous Phase: First 100 impacted (green) vs. non-impacted (purple) price paths, alongside resulting $z$-scores.}
    \label{fig:liquid}
\end{figure}

Figure \ref{fig:liquid} illustrates impact propagation within the Continuous Liquidity Phase. The system response decomposes into three temporal phases:
\begin{itemize}
    \item \textbf{Primary Impact:} Instantaneous price displacement caused by physical liquidity consumption of $Q$ shares.
    \item \textbf{Secondary Impact:} Intermediate propagation driven by Scalpers absorbing inventory imbalances and passing residual risk onto Liquidity Providers.
    \item \textbf{Impact Decay / Permanent Shift:} Relaxation dynamics determining whether the market absorbs the trade or enters a cascading regime.
\end{itemize}

We observe a clear transition from \textit{dissipative dynamics} at low volatility ($\sigma \lessapprox 4.21$ ticks, Figure \ref{fig:liquid:dissipative}), where impact decays over time, to \textit{non-dissipative dynamics} at high volatility ($\sigma \gtrapprox 4.21$ ticks, Figure \ref{fig:liquid:nondissipative}), where an initial trade triggers self-sustaining price cascades. The critical boundary ($\sigma \approx 4.21$ ticks, Figure \ref{fig:liquid:balanced}) exhibits perfectly permanent, non-decaying market impact.

\subsection{Market Impact in the Frozen Phase}
\begin{figure}[h!]
    \centering
    \begin{subfigure}[b]{0.45\textwidth}
        \includegraphics[width=\linewidth,height=0.15\textheight]{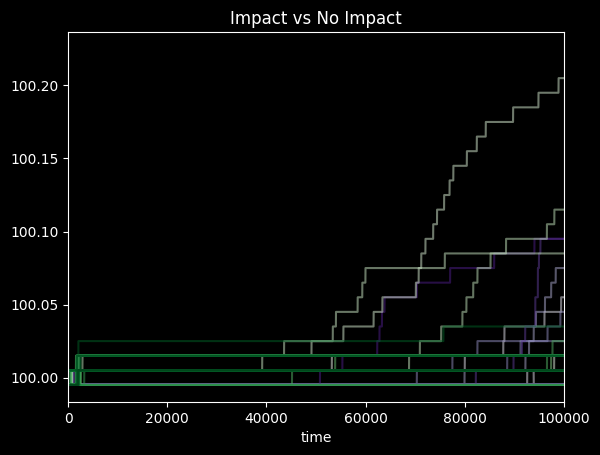}
        \includegraphics[width=\linewidth]{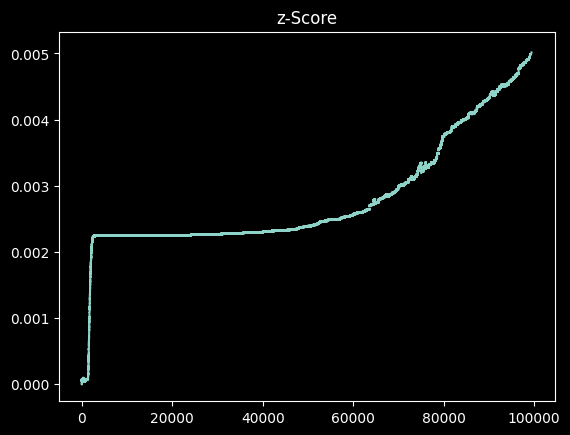}
        \caption{Cracking; $n=29$, $d=10$; $\sigma = 0.0051$ ticks}
        \label{fig:frozen:cracking}          
    \end{subfigure}
    \hfill
    \begin{subfigure}[b]{0.45\textwidth}
        \includegraphics[width=\linewidth,height=0.15\textheight]{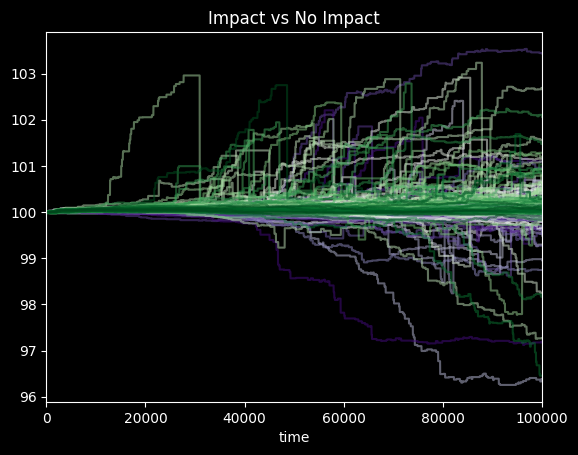}
        \includegraphics[width=\linewidth]{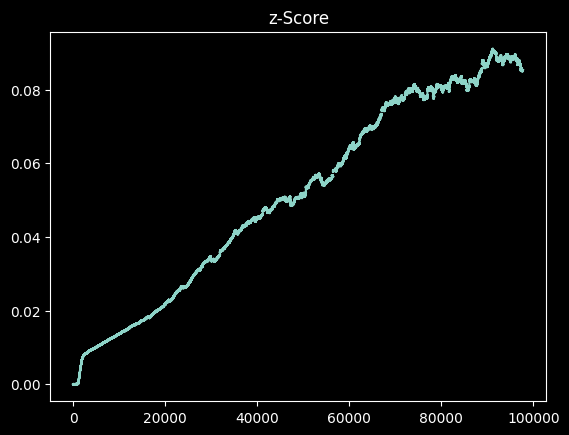}
        \caption{Melting; $n=14$, $d=10$; $\sigma = 0.432$ ticks}
        \label{fig:frozen:melting}          
    \end{subfigure}
    \caption{Impact and dissipation in the Frozen Phase: First 100 impacted (green) vs. non-impacted (purple) price paths, alongside resulting $z$-scores.}
    \label{fig:frozen}
\end{figure}

In the Frozen Phase (Figure \ref{fig:frozen}), impact dynamics display distinct solid-state mechanical analogs:
\begin{itemize}
    \item \textbf{Cracking Regime ($\sigma \lessapprox 0.1$ ticks):} Stress from the aggressive order dislodges a small subset of paths, while the majority remain locked (Figure \ref{fig:frozen:cracking}).
    \item \textbf{Melting Regime ($\sigma \gtrapprox 0.1$ ticks):} Impact energy destabilizes queue positions, causing widespread unfreezing and rapid structural transition across paths (Figure \ref{fig:frozen:melting}).
\end{itemize}

Crucially, these agentic impact profiles mirror empirical microstructure anomalies observed in modern equity and futures markets (see e.g., Bouchaud et al., Figure 4 \cite{bouchaud2003}).

\section{Discussion \& Implications}
The emergence of multi-agent reinforcement learning architectures in execution algorithms introduces non-linear feedback loops that are absent in classical, rule-based quantitative models. Our findings demonstrate that even minimal, homogeneous multi-agent LOB systems give rise to complex non-equilibrium thermodynamic behavior, regime shifts, and phase transitions.

These insights carry significant practical implications across quantitative finance:
\begin{itemize}
    \item \textbf{Execution Optimization \& Optimal Execution:} Traditional execution models based on square-root impact assumptions \cite{kyle1985} fail in non-dissipative agentic regimes. Execution algorithms must dynamically assess local market phase state ($n, d$) to prevent triggering strategic agentic cascades.
    \item \textbf{Systemic Risk \& Circuit Breakers:} The sharp first-order boundary separating continuous liquidity from order book collapse ($n < n_c$) illustrates how rapidly liquidity can evaporate. Regulators and exchanges must design adaptive circuit breakers that account for automated liquidity cancellation by dynamic agents during market stress.
    \item \textbf{Risk Management:} Standard Value-at-Risk (VaR) models assume continuous market clearing \cite{hasbrouck2007}. In agentic markets, transition into frozen or collapsed phases invalidates liquidity adjustments, necessitating stress-testing regimes built explicitly around multi-agent phase shifts.
\end{itemize}

An obvious limitation of the analysis in this paper is the homogeneity of the agent population. While this is helpful in terms of allowing us to map the entire phase space in terms of two parameters $(n,d)$ and summarize the phase behaviour, it is clearly unrealistic in terms of representing actual live trading LOBs. 

\section{Conclusion}
We have presented a statistical mechanics framework for Agentic Limit Order Books, establishing formal mappings for phase transitions, liquidity collapse boundaries, and non-dissipative market impact functions. Building on stochastic order book foundations \cite{cont2010}, we show that agentic interactions generate distinct macroscopic regimes — Collapsed, Continuous, and Frozen — each characterized by unique price discovery and impact profiles.

The main direct implication of this work is to with execution strategies, and we show that it is paramount to take into account the LOB phase behaviour when designing and running any such strategies, because the market response to any strategy can be markedly different in different phases.

Our main concern is with using the A-LOB as a simulation medium, in order to test, validate and risk assess any strategy, both before and after deploying it to a live environment. The results presented in this paper validate this approach, demonstrating that A-LOBs reproduce a number of realistic behaviours encountered in live trading.

\bibliographystyle{unsrt}

\end{document}